\documentclass[aps,preprint,prb,showkeys]{revtex4-1}
\usepackage{graphicx}
\usepackage{color}
\usepackage{multirow}

\begin{document}

\title{Tuning the Structural, Electronic, and Magnetic Properties of Germanene by the Adsorption of 3$d$ Transition Metal Atoms}

\author{Thaneshwor P. Kaloni}
\email{thaneshwor.kaloni@umanitoba.ca, +1-204-952-2900}
\affiliation{Department of Chemistry, University of Manitoba, Winnipeg, MB, R3T 2N2, Canada}

\begin{abstract}
The structural, electronic, and magnetic properties of 3$d$ transition metal (TM) atoms (Sc, Ti, V, Cr, Mn, Fe, Co, Ni, Cu, 
and Zn) adsorbed germanene are addressed using density functional theory. Based on the adsorption energy, TM atoms prefer 
to occupy at the hollow site for all the cases. The obtained values of the total magnetic moment vary from 0.97 $\mu_B$ to 
4.95 $\mu_B$ in case of Sc to Mn-adsorption, respectively. A gap of 74 meV with a strongly enhanced splitting of 67 meV 
is obtained in case of Sc-adsorption, whereas metallic states are obtained in case of Ti, Cr, Mn, Fe, and Co. Non-magnetic 
states are realized for Ni, Cu, and Zn-adsorption. Moreover, semiconducting nature is obtained for non-magnetic cases with a gap of 26 to 28 meV. Importantly, it is found that V-adsorbed germanene can host the quantum anomalous Hall 
effect. The obtained results demonstrate that TM atoms and nearest neighbour Ge atoms are ferro-magnetically 
ordered in the cases of V, Mn, Fe, Co, Ni, Cu, and Zn, while anti-ferromagnetic ordering is obtained for Sc, Ti, and 
Cr. In addition, the effects of the coverage of all TM atoms on the electronic structure and the ferro-magnetic and anti-ferro-magnetic coupling in case of Mn are examined. The results could help to understand the effect of TM atoms in a new class of two-dimensional materials beyond graphene and silicene.      
\end{abstract}
\maketitle

\section{Introduction}

Graphene is a material of interest particularly due to its unique electronic structure that features the Dirac cone. It has been proposed to be a great candidate for the nanoelectronic industry \cite{geim,castro}. In contrast to these expectations, the mass production and band gap opening in graphene are large hindrance. As consequence, its important to search for new classes of two-dimensional (2D) materials. In recent years, silicene and germanene have been expected to be counterparts of graphene \cite{ciraci}. Experimentally, the growth mechanism of silicene on metallic substrates and its probable impact in electronic devices have been addressed \cite{padova,Fleurence}. A strong interaction between the silicene and Ag metal by providing relatively large binding energy of about 460 meV has been found \cite{padova}, which is very large as compared to the binding energy of graphene on graphite (50 meV) and hence it would be difficult to exfoliate silicene from the Ag substrate. On the other hand, the sublattice symmetry of the silicene is broken due to the interaction of the silicene with ZrB$_2$ and hence a larger band gap of about 0.25 eV has been realized \cite{Fleurence}. Theoretically, the interaction of silicene with some semiconducting substrates (SiC and $h$-BN) has also been studied and it has been predicted that free-standing silicene can be realized on these substrates \cite{Liu,kaloni}. A smaller binding energy of about 57 meV \cite{kaloni} to 100 meV \cite{Liu} has been realized for silicene on $h$-BN and SiC substrates, which indicates that silicene can be exfoliated from these semiconducting substrates. As compared to C, Si, and Ge have larger ionic radii and as a result they promote $sp^3$ hybridization. The magnitude of the buckling in silicene and germanene is 0.46 \AA\ and 0.68 \AA, respectively, which is a result of the possible mixture of $sp^2$ and $sp^3$ hybridization in these materials. This buckling is responsible for the tunable band gap \cite{Ni,kaloni1,jpcc}. In addition, the intrinsic spin orbit interaction (SOI) of germanene is stronger than that for silicene and graphene. The magnitude of the SOI is found to be 46.3 meV, 4 meV, and 1 $\mu$eV, for germanene, silicene, and graphene, respectively \cite{Liu1}.

Recently, the growth of germanene on a GaAs(0001) substrate has been studied and it has been predicted that H passivated GaAs(0001) would be a potential way to realize quasi-free-standing germanene \cite{kaloni-jap}. A single layer of germanene without H passivation of the dangling Ga bonds is still strongly bound to the GaAs(0001), and as a result it would be difficult to exfoilate the germanene in order to make it free-standing. The fact is that after H intercalation between the interface of the germanene and GaAs(0001), it strongly reduces the interaction between germanene and the GaAs(0001) resulting in a binding energy of about 86 meV per Ge atom. In recent years, TM atoms doped graphene have been a topic of great interest. It has been found that TM atoms induce doping, scattering \cite{Geim3}, magnetism \cite{Foster}, and superconduction \cite{Uchoa} at the limit of dilute concentration and strongly modify the electronic structure at the limit of high concentration \cite{Giovannetti,udo}. Moreover, TM atom doped silicene has also been studied and it has been been demonstrated that 3$d$ TMs bind more strongly with silicene as compared to graphene and magnetism can be induced after doping in silicene \cite{zhang1}. Importantly, TM-adsorbed silicene shows the quantum anomalous Hall states, which is expected to be a potential candidate in spintronic devices \cite{kaloni-prb}. Furthermore, Be, Mg, and Ca-adsorbed silicene found to be a narrow gap semiconductor, while Ti and Cr-adsorbed silicene have been found to be half-metal. It is expected that the half-metallic ferromagnetic nature of Ti and Cr-adsorbed silicene could pave the way to build silicon-based spintronic devices \cite{peeters}. 

The structural, electronic, and magnetic properties of TM atoms adsorbed germanene have not been reported so far. However, from the 
electronic device applications point of view, it is important to understand the behaviour of TM atoms adsorbed germanene. To this 
aim, here we study Sc, Ti, V, Cr, Mn, Fe, Co, Ni, Cu, and Zn-adsorbed germanene in comparison to each other in the framework of the density functional theory. We find that Sc, Ti, V, Cr, Mn, Fe, and Co-adsorbed germanene induces magnetism, while non-magnetic semiconducting states are realized for Ni, Cu, and Zn. It is found that Sc, Ni, Cu, and Zn are semiconducting in nature with the largest gap of 74 meV for Sc-adsorption with a strong band splitting of 67 meV. Moreover, the effect of the coverage of TM atoms on the electronic structure has also been investigated. For the Mn doped case the ferro-magnetic and anti-ferro-magnetic coupling is examined in order to understand the coupling between two Mn atoms for variable distances. In addition, our results point out that V-adsorbed germanene can host the quantum anomalous Hall effect.

\section{Computational details}
All the calculations are performed using the generalized gradient approximation in the Perdew-Burke-Ernzerhof parametrization (PBE). \cite{pbe} 
The van der Waals interactions are taken into account in order to achieve the correct description of the long-range interaction and hence dispersions (PBE-D) \cite{grime}, which should be included in multilayers of 2D systems \cite{kaloni11,tp1,tp2,tp3}. A plane wave cutoff energy of 540 eV and a Monkhorst-Pack $16\times16\times1$ k-mesh are used. A $4\times4\times1$ supercell of germanene with a lattice constant of $a=16.24$ \AA\ and a vacuum layer of 15 \AA\ are used. The supercell contains 32 Ge atoms and 1 TM atom, which ensure that the density of the impurities is low enough in order to neglect the mutual interaction between TM atom impurities. The atomic positions are optimized until the forces have converged about to 0.005 eV/\AA. In addition, SOI is taken into account in all the calculations because germanene has stronger SOI of the order of 46.3 meV, which can not be neglected. It is well known that a finite onsite Coulomb interaction is essential to get an accurate description of atoms containing $d$ electrons \cite{udo,Wehling}, thus, we use an onsite Coulomb interaction of $U=4.0$ eV in our calculations.

\section{Germanene}

\begin{table} [ht]
\caption{The obtained values of structural and vibrational parameters for graphene, silicene, and germanene. The lattice constant, C$-$C, Si$-$Si, and Ge$-$Ge bond lengths, buckling, angle, band gap, cohesive energy, frequency of the E$_{2g}$-mode at the $\Gamma$-point, and Gr\"uneisen parameter.}
\begin{tabular}{|c|c|c|c|c|c|c|c|c|}
\hline
system  &$a$ (\AA)&$l$ (\AA)&$\Delta$ (\AA)&$\theta$ $(^{\circ})$&E$_{gap}$ (meV) &E$_{coh}$ (eV)/atom& E$_{2g}$ (cm$^{-1}$)&$\gamma_g$\\ 
\hline
\hline
graphene&2.46     &1.42     &0.00  &120&0.0 &8.12  &1598&1.80     \\
\hline
silicene&3.86     &2.26     &0.46  &116&1.6 &4.78  &558&1.65 \\
\hline
germanene&4.06    &2.44     &0.68  &112&24.3&3.09  &363&1.51   \\
\hline
\end{tabular}
\end{table}
\begin{figure}[ht]
\includegraphics[width=0.55\textwidth,clip]{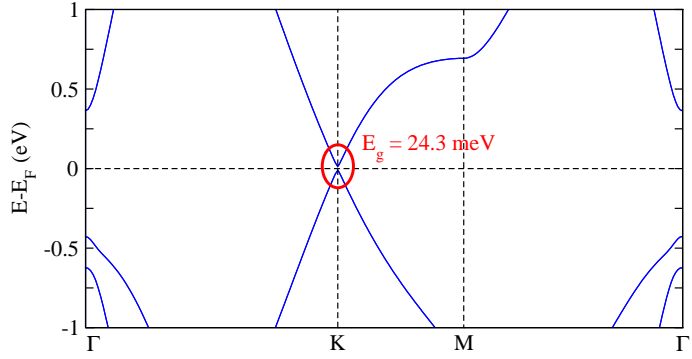}
\caption{The electronic structure of germanene by including SOI. A band gap of 24.3 meV is obtained (the portion is marked by a red circle).}
\end{figure}

In this section, for comparison, we present the data of the structural and vibrational properties of graphene, silicene, and 
germanene. Our calculated lattice parameters, nature of the materials (from a band structure point of view), and vibrational 
frequencies are presented in Table 1. Like graphene and silicene, germanene also is a bipartite lattice containing two 
interpenetrating triangular sublattices of Ge atoms. It is well know that the $\pi$ bonds between Ge atoms are weaker than those 
of Si and C atoms. However, the similarity between silicene and germanene is due to the fact that both of them 
are buckled structure unlike graphene. The buckling, which is defined as the perpendicular distance between two atoms in the unit 
cell, is 0.46 \AA\ in case of silicene \cite{peeters} and 0.68 \AA\ in case of germanene \cite{ciraci,kaloni-jap}. We obtain C$-$C, 
Si$-$Si, and Ge$-$Ge bond length of 1.42 \AA, 2.26 \AA, and 2.44 \AA, in a good agreement with the available reports. The obtained 
value of the cohesive energy decrease from graphene to germanene as expected, see Table 1. Graphene is metallic, while silicene and germanene are semiconducting with band gap of 1.6 meV and 24.3 meV, respectively \cite{Ni,kaloni1}. The electronic band structure of germanene using SOI is illustrated in Fig.\ 1. It is clear that $\pi$ bands are contributed from the $p_z$ orbitals and $\pi^*$ bands are due to the $p_z^*$ orbitals of the Ge atoms, in good agreement with previous findings \cite{ciraci,kaloni-jap}. In addition, calculated frequency of the E$_{2g}$-mode is 1598 cm$^{-1}$, 558 cm$^{-1}$, and 363 cm$^{-1}$ for graphene \cite{Zabel}, silicene \cite{Schaefer,kaloni-jap1}, and germanene \cite{Schaefer,kaloni-cpl}, respectively. Furthermore, our calculated values of the Gr\"uneisen parameter are found to be 1.80, 1.65, and 1.51 for graphene, silicene, and germanene, which in the good agreement with the previously reported values \cite{Zabel,kaloni-jap1,kaloni-cpl}.

\section{TM atoms adsorbed germanene}

\begin{figure}[ht]
\includegraphics[width=0.8\textwidth,clip]{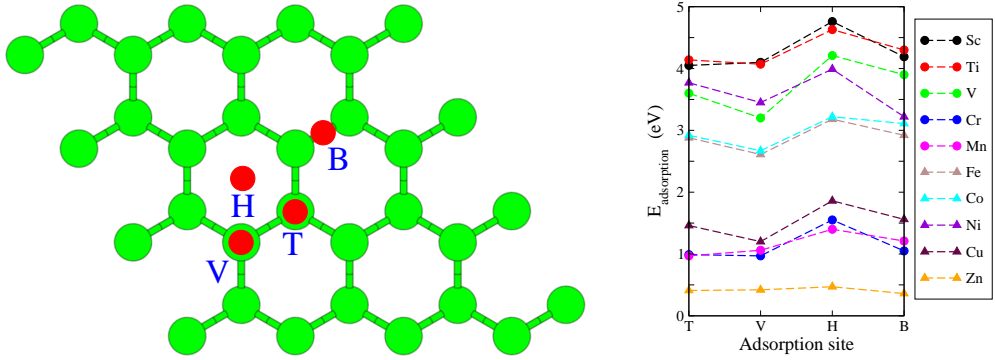}
\caption{Left side: Symbolic presentation of four possible adsorption sites indicated by top (T), valley (V), hollow (H), and 
bridge (B), where Ge and TM atoms are represented by green and red spheres. Right side: Absorption energy profile for 3$d$ TM atom 
doped germanene in four possible adsorption sites. The adsorption energy is calculated as $E_{\rm adsorption} = E_{\rm germanene} + E_{\rm TM} - E_{\rm germanene+TM}$, where $E_{germanene}$, $E_{TM}$, and $E_{germanene+TM}$ are the total energies for $4\times4\times1$ germanene supercell, total energy for an isolated TM atom, and the total energy for a $4\times4\times1$ germanene supercell with a single TM 
atom impurity, respectively.}
\end{figure}

\begin{table} [ht]
\caption{Dopant, the adsorption energy, the average distance of dopant form germanene sheet, Ge$-$Ge bond length, buckling, angle, total magnetic moment per supercell, magnetic moment contributed from $3d$ TM orbital, magnetic moment contributed from sum of all Ge atoms, nature of the bands (m$\rightarrow$metallic), and band splitting at the K point for TM atoms adsorbed on H-site.}
\begin{tabular}{|c|c|c|c|c|c|c|c|c|c|c|}
\hline
dopant &\begin{tabular}{c} E$_{adsorption}$ \\ (eV) \end{tabular} &\begin{tabular}{c} $d$ \\ (\AA) \end{tabular} &\begin{tabular}{c} Ge$-$Ge \\ (\AA) \end{tabular} &\begin{tabular}{c} $\Delta$ \\ (\AA) \end{tabular}&\begin{tabular}{c} $\theta$ \\ $(^{\circ})$ \end{tabular}&\begin{tabular}{c} total \\ ($\mu_B$) \end{tabular}&\begin{tabular}{c} $3d$ \\ ($\mu_B$) \end{tabular}&\begin{tabular}{c} Ge \\ ($\mu_B$) \end{tabular}&\begin{tabular}{c} E$_{gap}$  \\(meV) \end{tabular}&\begin{tabular}{c} E$_{sp}$ \\(meV) \end{tabular}\\
\hline 
\hline
Sc      &  4.62&1.21 &2.45-2.59   &0.62-0.74  &101-115    &0.97 &1.01&$-$0.04 & 74& 67  \\
\hline
Ti      &   4.46&0.71 &2.45-2.61   &0.64-0.76  &100-111    &1.13&1.99&$-$0.06  &m &--   \\
\hline
V      &    4.21&0.69 &2.45-2.71   &0.61-0.75  &95-111     &3.01&2.92&0.09   &1 & 19          \\
\hline
Cr     &    1.55&0.65 &2.45-2.63  &0.66-0.71  &95-111      &4.00&4.05&$-$0.05&m&21             \\
\hline
Mn     &    1.39&0.68 &2.45-2.67  &0.63-0.74   &97-112     &4.95&4.65&0.30   &m &11               \\
\hline
Fe     &    3.18&0.60 &2.46-2.62 &0.67-0.75    &100-112    &3.24&3.05&0.19 &m &31               \\
\hline
Co    &     3.22& 0.98 &2.48-2.81  &0.61-0.76   &78-111    &1.10&1.05&0.05 &m &--                \\
\hline
Ni    &     3.99&1.32 &2.44-2.45 &0.65-0.72    &111-112    &0.00  &0.00 &0.00&27 &26               \\
\hline
Cu    &     1.86&1.12 &2.44-2.47 &0.66-0.71   &112-114    &0.00 &0.00 &0.00 &29 &6               \\
\hline
Zn    &    0.99& 1.55 &2.44-2.46 &0.67-0.70     &112-113    &0.00 &0.00 &0.00&26 &24                   \\
\hline
\end{tabular}

\end{table}

The four possible adsorption sites for a single TM atom on germanene can be considered as on top of the upper Ge atoms (T-site), 
on top of the lower Ge atoms (V-site), above the center of the germanene hexagonal (H-site), and on top of the midway between 
two Ge atoms (B-site), see left side of Fig.\ 2. It is expected that TM atoms can occupy one of the sites mentioned above 
\cite{peeters}. We find that H-site is energetically most favourable in all the cases under study and hence, for the calculations 
of the electronic structures we only consider TM atoms adsorbed germanene at the H-site. We present a comparative energy diagram 
of all the four adsorption sites on the right side of Fig.\ 2. We find that the H-site adsorbed TM atoms have the largest adsorption 
energy as compared to adsorption on the another sites, see right side of Fig.\ 2, Table 2, and Table 3. A this point it is worth to mention that the adsorption energy of 3$d$ TM on graphene ranges from 1 eV to 2 eV \cite{ding}, while for silicene the value ranges from 2 eV to 3 eV \cite{kaloni-prb,peeters}. Therefore, the adsorption of TM atoms on germanene is highly favourable (except in case of Cu and Zn) in contrast to silicene and graphene, which indicates that germanene can be the best candidate to host TM atoms among all of the three 2D materials. For graphene, due to the planar structure which supports $sp^2$ hybridization, the $\pi$ orbitals are strongly coupled. However, in case of silicene, due to buckling and the resulting $sp^2$-$sp^3$ hybridization, the $\pi$ orbitals are coupled more weakly. As a result silicene has a larger adsorption energy as compared to graphene. Moreover, because of its larger buckling as compared to silicene, germanene has the weakest $\pi$ orbitals and hence more reactivity, as a result the highest binding energy is obtained.

\begin{table}[h]
\caption{Dopant, TM$-$Ge bond length, and adsorption energy for T-site and V-site, respectively.} 
\begin{tabular}{|c|c|c|c|c|}
\hline
dopant& \multicolumn{2}{|c|}{\multirow{1}{*}{TM$-$Ge (\AA)}}&\multicolumn{2}{|c|}{\multirow{1}{*}{E$_{adsorption}$ (eV)}}\\
\cline{2-5}
\hline
&T-site&V-site&T-site&V-site \\  
\hline
Sc &2.57&2.56&4.05&4.10  \\
\hline
Ti &2.55&2.54&4.14&4.07 \\
\hline
V &2.52&2.50&3.61 &3.20 \\
\hline
Cr &2.50&2.49&0.99&0.97 \\
\hline
Mn &2.48&2.47&0.97&1.06 \\
\hline
Fe &2.45&2.44&2.88&2.61 \\
\hline
Co &2.44&2.42&2.92&2.67 \\
\hline
Ni &2.42&2.41&3.77&3.45 \\
\hline
Cu &2.40&2.39&1.46&1.20\\
\hline
Zn &2.39&2.37&0.41&0.42\\
\hline
\end{tabular}
\end{table}

Our structural relaxation shows that a TM atom added on T-site and V-site replace the closest Ge atom by shifting the original 
Ge position. As a result a TM$-$Ge bond length of 2.57 \AA\ to 2.38 \AA\ is realized for Sc to  Zn in case of T-site and V-site, 
respectively, see Table 3. As expected, TM atom is connected to three nearest neighbour Ge atoms with slightly smaller bond length. This finding agrees well with the 3$d$ TM atoms adsorbed silicene \cite{kaloni-prb}. The TM atom is 
located at the midway between two Ge atoms by providing a slight distortion around the adsorption site in case of B-site 
adsorption. The TM atom occupies at the centre of the Ge hexagon in case of H-site adsorption. Around the adsorption site, 
we find a variable Ge$-$Ge bond lengths 2.45 \AA\ to 2.81 \AA\ for all the adsorbent, see Table 2. However, the Ge$-$Ge bond length is slightly modified as compared to pristine germanene with a value of 2.44 \AA\ \cite{peeters,ciraci,kaloni-cpl,kaloni-jap}. The buckling in the germanene sheet varies from 0.61 \AA\ to 0.76 \AA. The distance $d$, which is defined as the average height of TM atom from the germanene sheet is vary from 0.60 \AA\ to 1.55 \AA, see Table 2, which agrees well with TM atoms adsorbed silicene and graphene \cite{kaloni-prb,peeters,ding}. The obtained value of the angle (defined as the angle between Ge atom to next Ge atom to third Ge atom) ranges from 95$^{\circ}$ to 115$^{\circ}$, for Sc to Zn, respectively.

\section{Electronic structure}
\begin{figure}[ht]
\includegraphics[width=0.8\textwidth,clip]{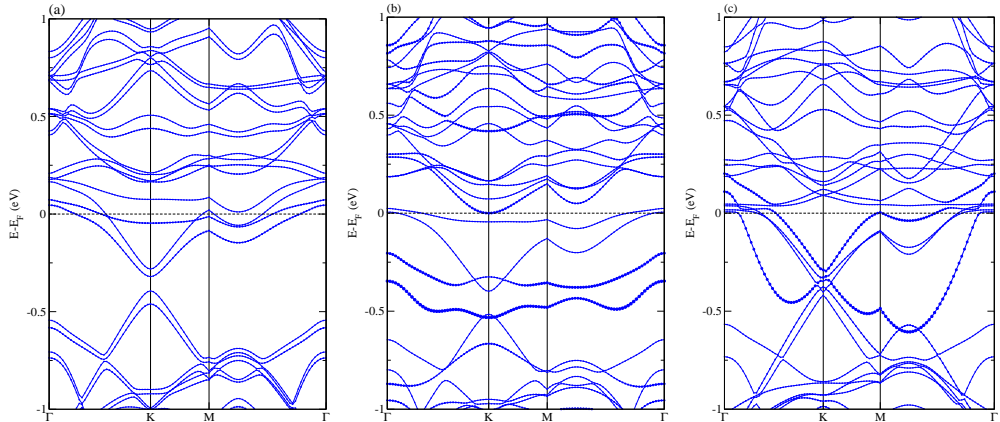}
\caption{Electronic structure (3$d$ TM states are represented by dots) for (a) Sc, (b) Ti, and (c) Cr-adsorbed germanene.} 
\end{figure}

In this section, we analyse the electronic structure of TM atoms adsorbed germanene at the energetically favourable H-site. 
Due to the charge redistribution between the 3$d$ state of the Sc and 4$p$ state of the Ge atoms, the Dirac like point of the 
germanene shifts below the Fermi level by 0.35 eV, see Fig.\ 3(a). The closer inspection of the partial partial density of states (PDOSs) of the $d$ orbital of Sc atom shows that the spin majority $d_{z^2}$, $d_{xy}$, and $d_{yz}$ as well as the spin minority $d_{3r^2}$ and $d_{yz}$ states contribute in the vicinity of the Fermi energy, see Fig.\ 4(a). The system becomes $n$-type doped with a gap of 74 meV; this gap could be shifted to the Fermi level such that system can be a perfect semiconductor \cite{kaloni-prb}. A stronger splitting of 67 meV is obtained, which agrees well with heavy atoms doped silicene \cite{kaloni-RRL}. Experimentally, it is expected that the materials having large band splitting can have great potential to construct spintronic devices \cite{PRL,amin}. The splitting is well distinguished from the splitting in pristine germanene, which is due to the SOI between the valence and conduction states. The double degeneracy of the valence and conduction bands is obtained by the SOI, which in fact break the time-reversal symmetry. Hence, the gap opening is the consequences of the sublattice symmetry breaking \cite{kaloni-RRL}. It is worth to mention that Sc-adsorbed silicene does not have splitting because silicene has a weaker SOI as compared to germanene \cite{zhang1}. Actually, there are two points which are interconnected. The band splitting is achieved due to the combined SOI of silicene/germanene and dopant. Normally, the SOI increases with increasing the size/atomic number of the atom. For example Au has the largest SOI among C, Si, Sc, Ge, and Au. In case of Sc-adsorbed silicene the combined SOI would be smaller than that of combined SOI of Au-adsorbed silicene, while combined SOI of Sc-adsorbed germanene would be definitely larger than that of Sc-adsorbed silicene because the strength of the SOI in germanene is 11 times higher than that in silicene. Note that the bands in the vicinity of the Fermi level are contributed from Ge orbitals with no contribution from Sc 3$d$ orbitals. A total magnetic moment of 0.97 $\mu_B$ per supercell is obtained. The main contribution of the magnetic moment comes Sc 3$d$ states of 1.01 $\mu_B$, whereas $-$0.04 $\mu_B$ comes from the sum of all 33 Ge atoms. The obtained value of the total magnetic moment shows a good agreement with the Sc-adsorbed silicene \cite{zhang1}. It should be noted that the host germanene sheet is significantly polarized by leaving anti-ferro-magnetic alignment between Sc and Ge atoms, which can be clearly understood by analysing the data presented in Table 2. 

\begin{figure}[ht]
\includegraphics[width=0.6\textwidth,clip]{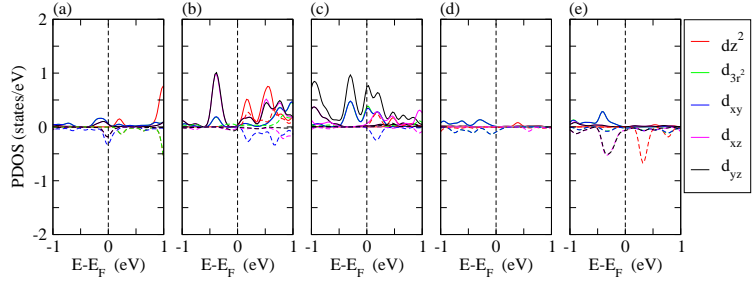}
\caption{The calculated partial density of states (PDOSs) of $3d$ orbitals for (a) Sc, (b) Ti, and (c) Cr, (d) Mn, and (e) Fe-adsorbed germanene. The solid and dashed lines represent the spin majority and minority, respectively.} 
\end{figure}

We obtain metallic nature of the bands for Ti-adsorbed germanene. It is clear that the bands in the Fermi level are contributed by Ge $p_z$ orbitals and bands at the energy range of about $-$0.24 eV to $-$0.35 eV are contributed by Ti 3$d$ orbitals, see Fig.\ 3(b). From the PDOSs of the Ti $3d$ orbitals, its is found that the spin majority $d_{z^2}$, $d_{xy}$, $d_{xz}$, and $d_{yz}$ orbitals contribute at the vicinity of the Fermi level together with the spin minority $d_{xy}$ and $d_{yz}$ orbitals, see Fig.\ 4(b). Here the gap in $p_z$ and $p_z^*$ is largely enhanced up to about 0.22 eV due to the interaction with the Ti atoms, which is responsible for breaking the sublattice symmetry as compared to pristine germanene. In this case the total 
magnetic moment of 1.93 $\mu_B$ per supercell is found, where Ti 3$d$ and Ge contributions amount to 1.99 $\mu_B$ and $-$0.06 $\mu_B$, respectively, see Table 2. The obtained value of the total magnetic moment agrees well with a recent report \cite{ciraci1}. 
The Ti-adsorbed germanene has been predicted to be an interesting system due to the fact that the magnetic 
moment and the nature of the material are significantly modified by a application of the perpendicular electric field. 
With certain values of the electric field the system behaves as half-metallic, which could be useful for 
spintronic devices \cite{ciraci1}. Like as Sc-adsorption, Ti and Ge atoms are also anti-ferro-magnetically ordered. 

\begin{figure}[h]
\includegraphics[width=0.8\textwidth,clip]{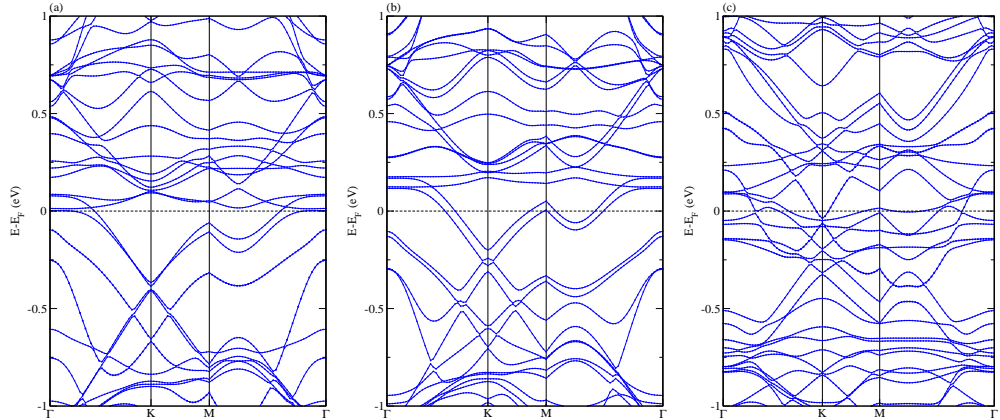}
\caption{Electronic structure for (a) Mn, (b) Fe, and (c) Co-adsorbed germanene.}
\end{figure}

The electronic band structure of Cr-adsorbed germanene is addressed in Fig.\ 3(c). We observe a strong hybridization between the 
Cr 3$d$ states to Ge 4$p$ states and as a result contributions of the 3$d$ states are achieved near the Fermi level. This implies that the system is metallic. The obtained data for the PDOSs of the Cr $3d$ orbitals show that the bands at the vicinity of the Fermi level are mainly contributed from the spin majority $d_{z^2}$, $d_{3r^2}$, $d_{xy}$, $d_{xz}$, and $d_{yz}$ orbitals, while a small contribution is coming from the spin minority $d_{xy}$ orbital, see Fig.\ 4(c). A band splitting of 21 meV is obtained, which is quite close to the band gap of pristine germenene. The total magnetic moment of 4.00 $\mu_B$ is obtained. The contribution of the magnetic moment from the $3d$ and the average of 33 Ge atoms are found to be 4.05 $\mu_B$ and $-$0.05 $\mu_B$, respectively. 
Our result for the magnetic moment agrees well with the Cr-adsorbed graphene and silicene \cite{park,kaloni-prb,peeters}. 
Recently, Cr-adsorbed graphene has been synthesized and the samples have been identified experimentally by analysing the shift in the Raman D and G peaks \cite{iqbal}, which could also be possible for Cr-adsorbed germanene. 

\begin{figure}[h]
\includegraphics[width=0.6\textwidth,clip]{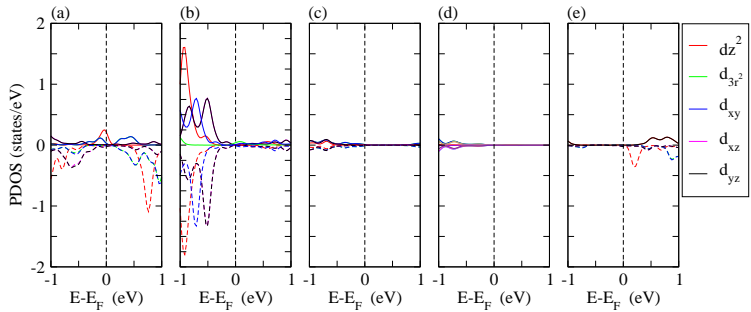}
\caption{The calculated PDOSs of 3$d$ states for (a) Co, (b) Ni, and (c) Cu, (d) Zn, and (e) V-adsorbed germanene. The solid and dashed lines represent the spin majority and minority, respectively.}   
\end{figure}

The electronic structure of Mn-adsorbed germanene is addressed in Fig.\ 5(a). The nature of the bands is similar to Cr-adsorption. However, in this case, Mn 3$d$ states are located far from the Fermi level, and the bands around the Fermi level are contributed from the Ge 4$p$ orbitals. The PDOSs of the Mn $3d$ orbitals addressed in Fig.\ 4(d) show that a minor contribution of the spin majority $d_{xy}$ and spin minority $d_{xy}$ orbitals are located at the vicinity of the Fermi level. A smaller band splitting of 11 meV is observed, which can be further enhanced by the application of an electric field \cite{kaloni1}. The total magnetic moment is found to be 4.95 $\mu_B$, while the Mn $d$ magnetic moment amounts to 4.65 $\mu_B$ and the Ge magnetic moment is found to be 0.30 $\mu_B$. This indicates that the Ge and the Mn atom are ferro-magnetically ordered. We also obtain the metallic nature of the bands in case of Fe and Co-adsorbed germanene, see Figs.\ 5(b-c). In case of Fe, the PDOSs of the Fe $3d$ orbitals show that the bands near to the Fermi level are contributed by the spin majority $d_{xy}$ and spin minority $d_{z^2}$, see Fig.\ 4(e). While for Co-adsorption, the spin majority $d_{z^2}$, $d_{3r^2}$, and $d_{xy}$ orbitals as well as $d_{z^2}$, $d_{3r^2}$, $d_{xy}$, $d_{xz}$, and $d_{yz}$ orbitals contribute the bands at the vicinity of the Fermi level, see Fig.\ 6(a). A band splitting of 31 meV is obtained in case of Fe-adsorption but we are not able to locate the band splitting in Co-adsorption due to strong hybridization (which causes the destruction in the bands) between Co 3$d$ and Ge 4$p$ orbitals. Total magnetic moment of 3.24 $\mu_B$ and 1.10 $\mu_B$ is obtained for Fe and Co-adsorption. The major contribution of the magnetic moment comes from TM $d$ orbitals (see Table 2) by keeping intact the ferro-magnetic ordering between TM and Ge atoms. The obtained value of the magnetic moment shows good agreement with TM atoms adsorbed silicene \cite{kaloni-prb,zhang1}.

\begin{figure}[h]
\includegraphics[width=0.8\textwidth,clip]{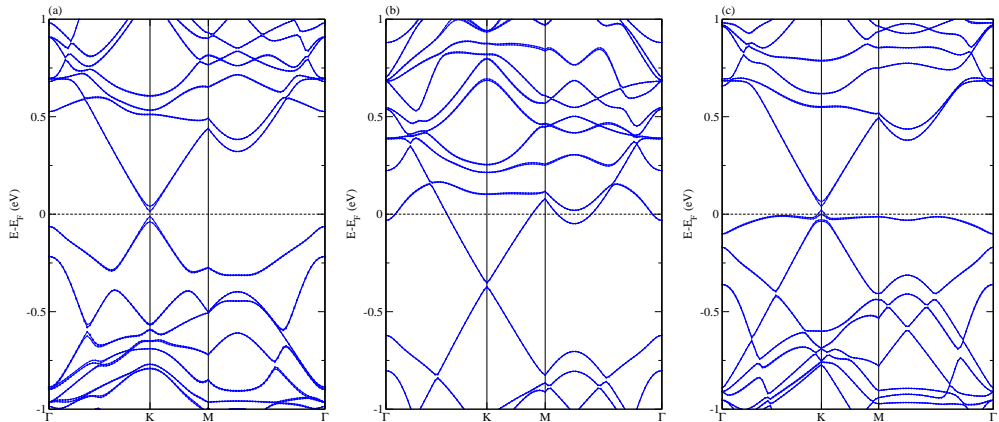}
\caption{Electronic structure for (a) Ni, (b) Cu, and (c) Zn-adsorbed germanene.}
\end{figure}

We obtain a perfect semiconducting nature of the electronic structure with a gap of 27 meV in case of Ni-adsorbed germanene, 
see Fig.\ 7(a). The contributions from the Ni 3$d$ states are located about 0.45 eV below the Fermi level, and they are coming from the spin majority and minority $d_{z^2}$, $d_{xy}$, and $d_{yz}$ orbitals, see Fig.\ 6(b). Furthermore, the obtained gap can be enhanced by applying an electric field \cite{Heinz}. Importantly, a band splitting of 26 meV is observed at the valence as well as conduction bands at the K-point. Such a splitting has been observed experimentally and theoretically for Pb adsorbed on the Ge(111) surface and it has been suggested that the materials having larger splitting can be utilized to develop spintronic devices \cite{Yaji}. As expected, the system is found to be non-magnetic, which agrees well with Ni-adsorbed silicene \cite{kaloni-prb}. The electrons from the 4$s^2$ orbital transfer to the 3$d^8$ and hence 3$d$ orbitals become completely filled and no magnetism is induced. We also realize the non-magnetic states in case of Cu-adsorption, see Fig.\ 7(b). In this case the Dirac like point of pristine germanene is shifted below the Fermi level (by leaving $n$-doped states) about 0.36 eV by opening a gap of 28 meV, slightly higher than that of pristine germanene of 24.3 meV. The $n$-type doping is obtained due to minute charge redistribution between Cu and nearest Ge atoms. Finally, we deal with Zn-adsorbed germanene, just like for graphene and silicene, in this case we also obtain non-magnetic states with a gap of about 26 meV close to the Fermi level, see Fig.\ 7(c). The fact is that the Zn atom is only weakly bound to the germanene and hence has minimal effect; as a result the opening of a gap is close to that of bare germanene. There is no charge redistribution between Ge and Zn atoms, which can be easily seen from Table 2. Moreover, the calculated PDOSs show that the contribution form Cu and Zn $d$ orbitals are located far from the Fermi level, see Fig.\ 6(c-d), which clearly indicates that the bands at the vicinity of the Fermi level are contributed from the Ge $p_z$ and $p_z^*$ orbitals.   
\begin{figure}[ht]
\includegraphics[width=0.7\textwidth,clip]{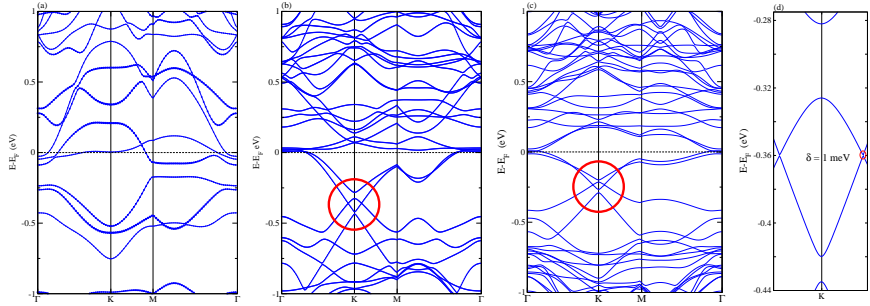}
\caption{Electronic structure of V-adsorbed germanene for (a) $2\times2\times1$, (b) $4\times4\times1$, (c) $5\times5\times1$ supercells, and (d) the zoomed portion of the red circle marked in (b).}
\end{figure}

In this section, we discuss an important part of the results, i. e. V-adsorbed germanene can host the quantum anomalous 
Hall effect. The quantum anomalous Hall effect arises because of the combined effect of the exchange-field (coming from the magnetic 
impurities, i.e. the V atom), which breaks the time-reversal symmetry, and the larger SOI (which induces band inversion and 
opens a nontrival gap). This effect has been predicted for TM atoms adsorbed topological insulators \cite{science}. Previously, 
the quantum anomalous Hall effect in graphene has been predicted \cite{Yang1,Niu2,Zhang2,Tse}. Very recently, 
the quantum anomalous Hall effect has been observed in graphene on the (111) surface of an antiferromagnetic insulator (BiFeO$_3$) \cite{MacDonald}. In addition, this effect has also been observed for silicene \cite{izawa,zhang1,kaloni-prb}. We find the 
quantum anomalous Hall effect only in case of V-adsorbed germanene; the combined effect of the exchange-field 
due to the V atom and SOI is sufficient to break the time-reversal symmetry and induce the band inversion, as a result the 
anomalous Hall effect is realized. Various coverages (11.11\%, 3.12\%, and 2.01\%) of the V atoms are calculated by considering $2\times2\times1$, $4\times4\times1$ and $5\times5\times1$ supercells of germanene. It it observed that only the supercells equal for or larger than $4\times4\times1$ are able to induce the quantum anomalous Hall effect. This indicates that the interaction between the V atom and its periodic image is low enough such that the quantum anomalous Hall effect can not be suppressed \cite{kaloni-prb}. We calculate the electronic structure of the V-adsorbed germanene with the coverages mentioned above, see Fig.\ 8. In Fig.\ 8(a), the band structure of a $2\times2\times1$ supercell of germanene with one V atom (concentration of the V atom is 11.11\%) is calculated. The distance between the V atom and its periodic images is 8.12 \AA. In this case the quantum anomalous Hall effect is not observed due to the strong interaction between the V atoms, which probably destroy the effect. The bands near the Fermi level contributions from both the V $3d$ orbitals and Ge $4p_z$ orbitals, see Fig.\ 8(a). 

However, in case of smaller coverages of the V atoms ($4\times4\times1$ and $5\times5\times1$ supercells) the bands near the Fermi level are not modified strongly and the bands are essentially coming from the Ge 4$p_z$ orbitals, see Fig.\ 8(b-c). The calculated PDOSs for the V $3d$ orbitals show that the bands close to the Fermi level are contributed by the spin majority $d_{yz}$ and spin minority $d_{z^2}$, $d_{3r^2}$, and $d_{xy}$ orbitals, see Fig.\ 6(e). The distance between the V atom and its periodic image is large enough at 16.24 \AA\ and 20.30 \AA, respectively for $4\times4\times1$ and $5\times5\times1$ supercells to avoid the interaction between the V atoms. Hence, the quantum anomalous Hall effect is realized in the latter two cases. The quantum anomalous Hall states are located below the Fermi level by about 0.3 eV and 0.25 eV, which is indicated by the red circle. The zoomed portion of the red circle in Fig.\ 8(b) is presented in Fig.\ 8(d). We find a nontrivial gap of $\delta=1$ meV, which agrees well with the recently calculated value of the $\delta$ for graphene on BiFeO$_3$ \cite{MacDonald}.

\section{Effect of the coverage}

\begin{figure}[h]
\includegraphics[width=0.6\textwidth,clip]{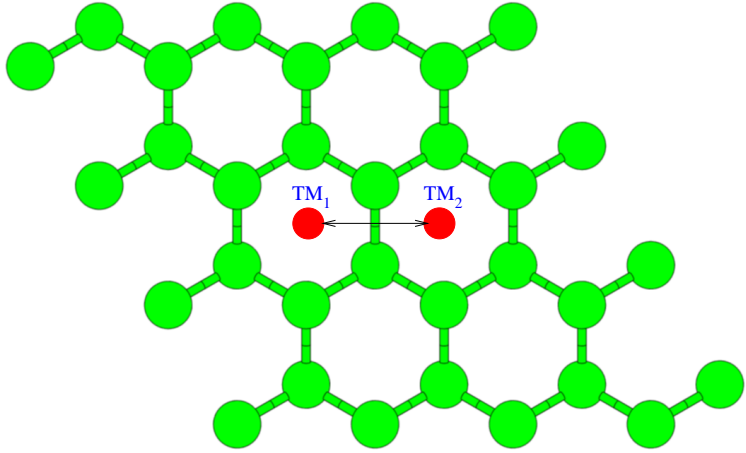}
\caption{Symbolic presentation of the adsorption of two TM atoms on H-site, where Ge and TM atoms are represented by green and
red spheres. A double sided arrow shows the distance of 4.06 \AA\ between TM$_1$ and TM$_2$ atoms.} 
\end{figure}

The structure under consideration for two TM atoms (the concentration of TM atom is about 6.25\%) adsorbed germanene on H-site is depicted in Fig.\ 9. The distance between TM$_1$ and TM$_2$ amounts to be $\sim$4.06 \AA, which is equivalent to the lattice constant of a $1\times1\times1$ unit cell for germanene. In case of two Sc-adsorbed germanene, a strong distortion in the structure is observed as compared to single Sc-adsorbed germanene. The symmetry is broken due to the distortion, as a result a large direct/indirect band gap of 290/220 meV is achieved, see Fig.\ 10(a). The obtained value of the band gap is $\sim12$ times larger than that of pristine germanene, which could be interesting for germanene based electronic devices \cite{PRL}. The bands close to the Fermi level are contributed from Ge $p_z$ orbitals, it has been demonstrated that the bands near the Fermi level are mainly not contributed from TM atom if the concentration of TM atom is low ($\sim$16 \%) \cite{arxiv13}. A band gap of 40 meV is obtained for two Ti-adsorbed germanene, see Fig.\ 10(b), while single Ti-adsorbed germanene is metallic in nature as discussed in previous section. In case of two Cr-adsorption, the Dirac like cone shifted above the Fermi level (120 meV) by leaving the $p$-doped states with a band gap of 11 meV at the $\pi$ and $\pi^*$ orbitals of Ge atoms, see Fig.\ 10(c). Whereas, the metallic states are obtained for two Mn, Fe, and Co-adsorbed germanene (see Figs.\ 10(d-f)), the bands are qualitatively similar to a single Mn, Fe, and Co-adsorption. Moreover, the calculated electronic band structure of two Ni-adsorbed germanene is addressed in Fig.\ 10(g), a direct/indirect band gap of 63/172 meV is obtained. In case of two Cu-adsorption, the Dirac like cone perturbed and shifted below the Fermi level (420 meV) by inducing a gap of 21 meV, see Fig.\ 10(h), this observation is very similar to single Cu-adsorption. Finally, the electronic band structure of two Zn-adsorbed germanene is addressed in Fig.\ 10(i). In contract to single Zn-adsorption, the system is found to be perfect semiconductor with a band gap of 25 meV. In all the cases, the electronic band structures are modified, the modification can be attributed by the strong interaction between two TM impurities and their images, this fact has also been observed for TM atoms adsorbed silicene \cite{kaloni-prb}.

\begin{figure}[ht]
\includegraphics[width=0.8\textwidth,clip]{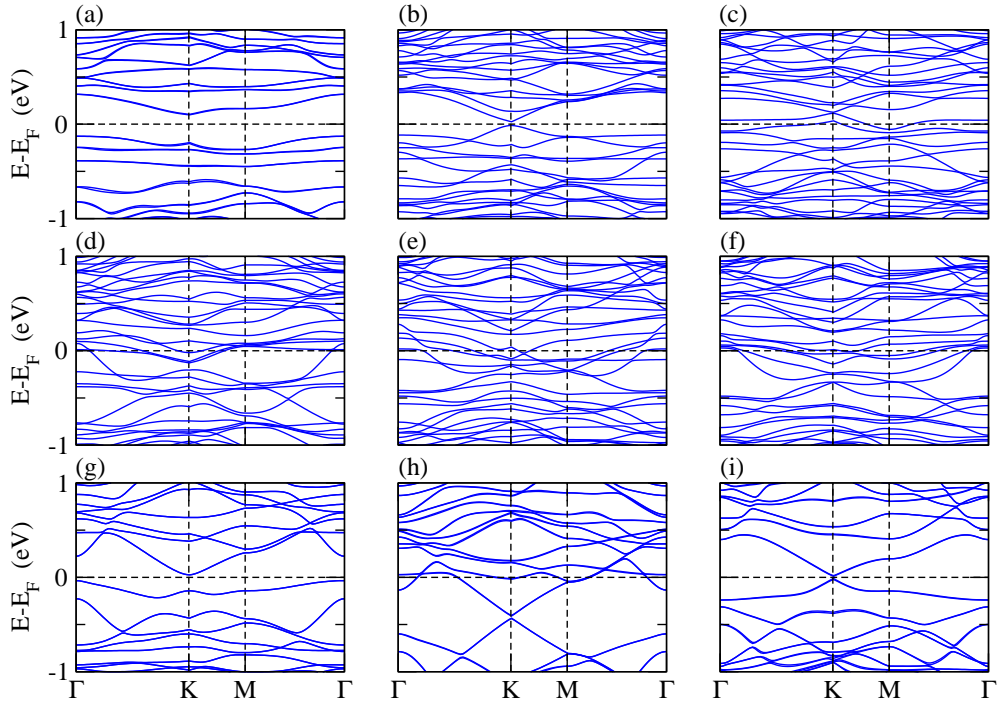}
\caption{The electronic band structure for TM atoms coverage of 6.25\% for (a) Sc, (b) Ti, and (c) Cr, (d) Mn, (e) Fe, (f) Co, (g) Ni, (h) Cu, and (i) Zn-adsorbed germanene.} 
\end{figure}

\section{Ferro/anti-ferro-magnetic coupling for Mn-adsorbed germanene}
\begin{figure}[h]
\includegraphics[width=0.3\textwidth,clip]{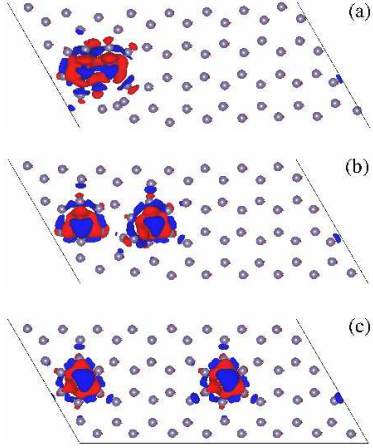}
\caption{The charge density difference isosurface for two Mn-adsorbed $8\times4\times1$ supercell of germanene for (a)-(b) anti-ferro-magnetic coupling and (c) ferro-magnetic coupling between Mn$_1$ and Mn$_2$ atoms. The isosurface corresponds to a value of $2\cdot 10^{-3}$ electrons/\AA$^{-3}$.} 
\end{figure}

In this section, we address the ferro/anti-ferromagnetic coupling between two Mn (Mn$_1$ and Mn$_2$) atoms. Here, we investigate two Mn-adsorbed $8\times4\times1$ supercell of germanene, the supercell is fairly adequate in order to model such a systems \cite{wu3}. This supercell contains 64 Ge and 2 Mn atoms with a lattice parameters of $a=30.95$ \AA, $b=15.48$ \AA, and a vacuum layer of 15 \AA\ on top. There are three possible spin configurations for Mn atoms, categorized as (i) ferro-magnetic (parallel spin), (ii) anti-ferro-magnetic (anti-parallel spin), and paramagnetic (both Mn atoms are far from each other such that they do not interact). The paramagnetic situation can be observed if the energy difference between parallel and anti-parallel configurations is very small (of the order of few meV) \cite{wu3}. In order to calculate this situation, one has to consider very large supercell, due to the computational limitation we do not considered paramagnetic situations in the calculations. We consider three configurations such that the distance between Mn$_1$ and Mn$_2$ atoms set to be (a) 2.98 \AA, (b) 7.69 \AA, and (c) 15.47 \AA\ (see Fig.\ 11) and systems are fully relaxed thereafter. The energy difference between the anti-ferro-magnetic and ferro-magnetic coupling is found to be $-0.33$ eV, $-$0.16 eV, and 0.8 eV for the configurations (a), (b), and (c), respectively. Which indicates that the anti-ferro-magnetic coupling is energetically favorable for the configurations (a) and (b), while the ferro-magnetic coupling is energetically favorable for configuration (c). In case of configuration (a), germanene sheet is strongly modified due to the strong interaction between two Mn impurities, see Fig.\ 11(a). Whereas, a small modification is observed for the configurations (b) and (c) because of the larger distance between two Mn impurities (Figs.\ 11(b)-(c)), which imply that the interaction between Mn atoms is suppressed. In case of the configurations (a) and (b), the total magnetic moment is found to be 0.0 $\mu_B$, which reflects the anti-ferro-magnetic coupling between Mn$_1$ and Mn$_2$ atoms, see the calculated charge density isosurface. However, in case of configuration (c) the total magnetic moment is found to be double than that of single Mn-adsorbed germanene, indicating the ferro-magnetic coupling, see Fig.\ 11(c). For configuration (c), see Fig.\ 11(c), the isosurface for both Mn$_1$ and Mn$_2$ atoms is exactly same, which is not a case for anti-ferro-magnetic coupling, compare Figs.\ 11(a)-(b). Thus, anti-ferro-magnetic coupling is energetically favorable if the distance between two Mn atoms is about 7.69 \AA, while ferro-magnetic coupling seems to be energetically favorable if two Mn atoms are separated by a large distance of about 15.47 \AA. These findings are quantitatively matching with Mn doped graphene \cite{wu3}.

\section{Conclusion}
In conclusion, based on the first-principles calculations, we report the structural, electronic, and magnetic properties of 
3$d$ TM atoms, such as Sc, Ti, V, Cr, Mn, Fe, Co, Ni, Cu, and Zn-adsorbed germanene. Based on the adsorption energy, we found that 
the adsorption of a TM atom at the hollow site is energetically most favourable for all the cases. The obtained value of the total 
magnetic moment ranges from 0.97 $\mu_B$ to 4.95 $\mu_B$ for Sc to Mn-adsorbed germanene, respectively. A gap of 
74 meV with a strongly enhanced splitting of 67 meV is found for Sc-adsorption, while metallic states are obtained for Ti to Co-adsorption. Non-magnetic states are realized in case of Ni, Cu, and Zn. Furthermore, a semiconducting nature is obtained for non-magnetic cases while a gap of 26 to 28 meV. Our result also point out that V-adsorbed germanene could host the quantum anomalous Hall effect, which is in agreement with V-adsorbed silicene. Our results show that TM and nearest neighbour Ge atoms are ferro-magnetically ordered in case of V, Mn, Fe, Co, Ni, Cu, and Zn, whereas an anti-ferromagnetic ordering is achieved for Sc, Ti, and Cr-adsorbed germanene. Furthermore, the effects of the coverage of TM atoms on the electronic structure for all TM atoms under consideration and the ferro-magnetic as well as anti-ferro-magnetic coupling in case of Mn-doped germanene are studied. These results contribute to the understanding of the effect of TM atoms impurities in the electronic structure of a new class of 2D material beyond graphene and silicene. 
 
\section*{ACKNOLEDGEMENT}
It is my great pleasure to acknowledge G. Schreckenbach and M. S. Freund for fruitful discussions.

\end{document}